\documentclass[12pt]{article}
\usepackage{graphics}

\def\hybrid{\topmargin -20pt    \oddsidemargin 0pt
        \headheight 0pt \headsep 0pt
        \textwidth 6.25in       
        \textheight 9.5in       
        \marginparwidth .875in
        \parskip 5pt plus 1pt   \jot = 1.5ex}

\hybrid

\def\baselinestretch{1.2}

\catcode`\@=11

\def\marginnote#1{}
\newcount\hour
\newcount\minute
\newtoks\amorpm
\hour=\time\divide\hour by60
\minute=\time{\multiply\hour by60 \global\advance\minute by-\hour}
\edef\standardtime{{\ifnum\hour<12 \global\amorpm={am}%
        \else\global\amorpm={pm}\advance\hour by-12 \fi
        \ifnum\hour=0 \hour=12 \fi
        \number\hour:\ifnum\minute<10 0\fi\number\minute\the\amorpm}}
\edef\militarytime{\number\hour:\ifnum\minute<10 0\fi\number\minute}

\def\draftlabel#1{{\@bsphack\if@filesw {\let\thepage\relax
   \xdef\@gtempa{\write\@auxout{\string
      \newlabel{#1}{{\@currentlabel}{\thepage}}}}}\@gtempa
   \if@nobreak \ifvmode\nobreak\fi\fi\fi\@esphack}
        \gdef\@eqnlabel{#1}}
\def\@eqnlabel{}
\def\@vacuum{}
\def\draftmarginnote#1{\marginpar{\raggedright\scriptsize\tt#1}}

\def\draft{\oddsidemargin -.5truein
        \def\@oddfoot{\sl preliminary draft \hfil
        \rm\thepage\hfil\sl\today\quad\militarytime}
        \let\@evenfoot\@oddfoot \overfullrule 3pt
        \let\label=\draftlabel
        \let\marginnote=\draftmarginnote
   \def\@eqnnum{(\theequation)\rlap{\kern\marginparsep\tt\@eqnlabel}%
\global\let\@eqnlabel\@vacuum}  }

\def\preprint{\twocolumn\sloppy\flushbottom\parindent 2em
        \leftmargini 2em\leftmarginv .5em\leftmarginvi .5em
        \oddsidemargin -.5in    \evensidemargin -.5in
        \columnsep .4in \footheight 0pt
        \textwidth 10.in        \topmargin  -.4in
        \headheight 12pt \topskip .4in
        \textheight 6.9in \footskip 0pt
        \def\@oddhead{\thepage\hfil\addtocounter{page}{1}\thepage}
        \let\@evenhead\@oddhead \def\@oddfoot{} \def\@evenfoot{} }

\def\numberbysection{\@addtoreset{equation}{section}
        \def\theequation{\thesection.\arabic{equation}}}

\def\underline#1{\relax\ifmmode\@@underline#1\else
        $\@@underline{\hbox{#1}}$\relax\fi}

\def\titlepage{\@restonecolfalse\if@twocolumn\@restonecoltrue\onecolumn
     \else \newpage \fi \thispagestyle{empty}\c@page\z@
        \def\thefootnote{\fnsymbol{footnote}} }

\def\endtitlepage{\if@restonecol\twocolumn \else \newpage \fi
        \def\thefootnote{\arabic{footnote}}
        \setcounter{footnote}{0}}  

\catcode`@=12
\relax

\def\figcap{\section*{Figure Captions\markboth
        {FIGURECAPTIONS}{FIGURECAPTIONS}}\list
        {Figure \arabic{enumi}:\hfill}{\settowidth\labelwidth{Figure
999:}
        \leftmargin\labelwidth
        \advance\leftmargin\labelsep\usecounter{enumi}}}
 \relax
\def\tablecap{\section*{Table Captions\markboth
        {TABLECAPTIONS}{TABLECAPTIONS}}\list
        {Table \arabic{enumi}:\hfill}{\settowidth\labelwidth{Table
999:}
        \leftmargin\labelwidth
        \advance\leftmargin\labelsep\usecounter{enumi}}}
 \relax
\def\reflist{\section*{References\markboth
        {REFLIST}{REFLIST}}\list
        {[\arabic{enumi}]\hfill}{\settowidth\labelwidth{[999]}
        \leftmargin\labelwidth
        \advance\leftmargin\labelsep\usecounter{enumi}}}
 \relax
\makeatletter
\newcounter{pubctr}
\def\publist{\@ifnextchar[{\@publist}{\@@publist}}
\def\@publist[#1]{\list
        {[\arabic{pubctr}]\hfill}{\settowidth\labelwidth{[999]}
        \leftmargin\labelwidth
        \advance\leftmargin\labelsep
        \@nmbrlisttrue\def\@listctr{pubctr}
        \setcounter{pubctr}{#1}\addtocounter{pubctr}{-1}}}
\def\@@publist{\list
        {[\arabic{pubctr}]\hfill}{\settowidth\labelwidth{[999]}
        \leftmargin\labelwidth
        \advance\leftmargin\labelsep
        \@nmbrlisttrue\def\@listctr{pubctr}}}
 \relax
\makeatother
\newskip\humongous \humongous=0pt plus 1000pt minus 1000pt

\newif\ifdtup

\relax

\def\be{\begin{equation}}
\def\ee{\end{equation}}
\def\ba{\begin{eqnarray}}
\def\ea{\end{eqnarray}}

\def\del{\partial}

\def\e{\epsilon}

\def\m{\mu}
\def\n{\nu}
\def\om{\omega}

\def\l{\lambda}
\def\L{\Lambda}

\def\no{\noindent}

\def\qq{\qquad}

\def\IR{\relax{\rm I\kern-.18em R}}

\def \ha {{1\over 2}}

\def \ov {\over}

\def\const{{\rm const.}}

\def\IR{\relax{\rm I\kern-.18em R}}
\def\inv{^{\raise.15ex\hbox{${\scriptscriptstyle -}$}\kern-.05em 1}}

\def\tL{{\tilde L}}

\def\crbig{\\\noalign{\vspace{3mm}}}

\begin{document}

\renewcommand{\theequation}{\thesection.\arabic{equation}}

\newcommand{\beq}{\begin{equation}}
\newcommand{\eeq}[1]{\label{#1}\end{equation}}
\newcommand{\ber}{\begin{eqnarray}}
\newcommand{\eer}[1]{\label{#1}\end{eqnarray}}
\newcommand{\eqn}[1]{(\ref{#1})}
\begin{titlepage}
\begin{center}

\hfill NEIP-00-015\\
\vskip -.05 cm
\hfill hep--th/0008025\\
\vskip -.05 cm
\hfill July 2000\\

\vskip .7in

{\large \bf Supersymmetric phases of finite-temperature strings II}

\vskip 0.5in

{\bf Konstadinos Sfetsos}
\vskip 0.1in
{\em Institut de Physique, Universit\'e de Neuch\^atel\\
     Breguet 1, CH-2000 Neuch\^atel, Switzerland}\\
{\tt sfetsos@mail.cern.ch}
\vskip .2in

\end{center}

\vskip .7in

\centerline{\bf Abstract}

\no
It was recently proposed that there exist stable supersymmetric phases 
for finite temperature superstings. This issue was investigated
using an effective supergravity which takes into account 
massive winding modes. Such a theory admits BPS solutions that do not 
suffer from Hagedorn-type instabilities.
We extend several aspects of this work. 
First we restrict to the real-field sector of the theory and 
allow, in general, for unequal right and left ($\om_+$ and $\om_-$) 
winding fields.
Then, by further specializing to type-II theories (IIA, IIB and a
self-dual hybrid) we construct the most general 1/2-BPS solution 
and reveal several
new features arising in various consistent truncations 
($\om_-=\pm \om_+$, $\om_-\gg \om_+$).
In the heterotic case we investigate the general properties of the solution
which is presented in a closed form in the limit of infinitely large 
left-winding field ($\om_-\gg \om_+$).

\vskip .4 cm
\noindent
\end{titlepage}
\vfill
\eject

\def\baselinestretch{1.2}
\baselineskip 16 pt
\noindent

\def\tT{{\tilde T}}
\def\tg{{\tilde g}}
\def\tL{{\tilde L}}


\section{Introduction and generalities}

Studying string theory at finite temperature is important 
due to the relevance for the physics of the very early universe.
It is also 
important for the further understanding of string theory itself,
since the interest effects, such as Hagedorn-type instabilities,
are due to massive windings string states becoming massless as
the temperature is raised. These and related issues
were investigated in the past in a series of works, notably 
in \cite{hag}-\cite{neil}.
In these works, the interactions of massive string states 
among themselves and with the massless states 
were either ignored or assumed that they would not changed drastically the 
physical picture. As far as supersymmetry is concerned, it was always 
assumed that it necessarily breaks once the temperature is turned on. 

The construction of an effective supergravity for finite temperature
$N=4$ superstrings in five-dimensions that takes into account the relevant 
massive windings states was done for the heterotic and type-II
cases in \cite{ak} and extended in a U-duality invariant way,
that encompasses all perturbative string theories, in \cite{adk}.
These theories are effectively 
four-dimensional and have as a vacuum solution flat 
space with zero winding fields and constant moduli, which breaks supersymmetry.
Furthermore, the more general theory of \cite{adk}, 
reproduces all known Hagedorn-type instabilities for each 
perturbative string theory separately.
 
It was only recently that, non-trivial solutions to these theories were
found preserving 1/2 of supersymmetries and 
not suffering from Hagedorn-type instabilities \cite{BBDS}. 
Their very existence
suggests that when superstrings are heated 
up from zero temperature, they might prefer to go into this more symmetric
phase which is stable due to supersymmetry, and that the ``ordinary" 
high-temperature instabilities do not occur, i.e., there is no
Hagedorn temperature. It is the purpose of this paper to further pursue 
this idea and generalize several aspects of the work of \cite{BBDS}.

Let us briefly review the most relevant aspects of our effective 
theory. For details on its construction the reader should consult the 
original literature \cite{ak,adk} and for a shorter review suitable for our
purposes and further physical motivation, the ref.  \cite{BBDS}. 
Our effective theory is a four-dimensional $N=1$ supergravity theory 
coupled to 9 chiral 
multiplets corresponding to the 3 complex moduli fields $S$, $T$ and $U$ 
and the 6 complex winding fields $Z_A^\pm$, $A=1,2,3$. 
As any generic four-dimensional $N=1$ supergravity
theory it is characterized 
by a K\"ahler potential $K$ and a holomorphic superpotential $W$, but 
we will not present their explicit expressions here.
It was shown in \cite{BBDS} that for any four-dimensional $N=1$ supergravity
theory coupled to chiral multiplets, one may consistently restrict 
to the real part of the bosonic fields and set the imaginary part to 
zero. In particular, this procedure leads to 1/2-BPS solutions in a natural 
way, after examining the Killing spinor equations of the theory.
Furthermore, for the theory of \cite{adk} 
describing finite temperature superstrings in a U-duality invariant way,
we focused on the real directions defined by 
${\rm Im}\, Z_A^{\pm} = {\rm Im}\, S = 
{\rm Im}\, T = {\rm Im}\, U = {\rm Re }\, (Z_A^+-Z_A^-)=0$ which, in total,
represent 6 real fields \cite{BBDS}. This consistent truncation lead to 
important simplifications and made possible to find the general 1/2-BPS 
solution
for the type-II cases and systematically investigate the heterotic case as
well.

In this paper we will relax the last of the above conditions 
that equates the right  
winding field $Z^+_A$ with the left winding field $Z_A^-$, 
and impose only the condition that all fields are real.
Hence, we remain with 9 real fields in total which we rename as
\be
s = {\rm Re}S\ , \qq t={\rm Re}T\ ,\qq  u={\rm Re}U\ , \qq
z_A^\pm = {\rm Re} Z_A^\pm\ ,\quad A=1,2,3\ .
\label{condiii}
\ee
We will see that this generalization reveals some new features of the theory.

Let us introduce for convenience the quantities\footnote{We 
follow the notation of \cite{BBDS} 
appropriately modified when necessary.}
\be
x_\pm^2 = \sum_{A=1}^{3} (z_A^\pm)^2 ~,\qq H_A^\pm = {z_A^\pm 
\over 1 - x_\pm^2}\ .
\label{hreal} 
\ee
It is also helpful to trade $s$, $t$ and $u$ for  
$\phi_1$, $\phi_2$ and $\phi_3$ as follows:
\be
s=e^{-2 \phi_1} \ , \qq t=e^{-2\phi_2} \ ,\qq  u=e^{-2\phi_3} \ .
\ee
With these definitions, we obtain a simplified form of the 
effective supergravity with bosonic Lagrangian density
\be
{1\ov \sqrt{g}} {\cal L} = {1 \over 4}R -{1 \over 2} 
\sum_{i=1}^3 (\partial_\mu\phi_i)^2
- \sum_{A=1}^{3} {(\partial_\mu z^+_A)^2  \over (1 - x_+^2)^2} 
- \sum_{A=1}^{3} {(\partial_\mu z^-_A)^2 \over (1 - x_-^2)^2} 
- V ~,  
\label{laagra}
\ee
where the scalar potential is given in terms of a superpotential $W$ as
\be
V={1 \over 4} \sum_{i=1}^3 \left({\partial W \over 
\partial \phi_i}\right)^2 + {1 \over 8} \sum_{A=1}^3 (1 - x_+^2)^2 
\left({\partial W \over \partial z^+_A}\right)^2
+ {1 \over 8} \sum_{A=1}^3 (1 - x_-^2)^2 
\left({\partial W \over \partial z^-_A}\right)^2 -
{3\ov 4} W^2 \ ,
\label{dina1}
\ee
with
\be 
\begin{array}{rcl}
W &=& {1 \over 2} 
e^{\phi_1+\phi_2+\phi_3} - 2 e^{\phi_1} {\rm sinh} (\phi_2+\phi_3) 
H_1^+ H_1^- +e^{-\phi_1}\left(e^{\phi_2-\phi_3} H_2^+ H_2^- 
+ e^{\phi_3-\phi_2} H_3^+ H_3^-
\right) \crbig
&=& {1\over\sqrt{stu}}\left( {1\over2} + (tu-1)H_1^+ H_1^- + su H_2^+ H_2^- 
+ stH_3^+ H_3^- 
\right)\,.
\end{array}
\label{supereal} 
\ee 

In this paper we will use units where the four-dimensional gravitational 
coupling $\kappa$ has been normalized to $\sqrt{2}$.

We make the domain wall ansatz (see, for instance, \cite{cvso})
for the metric that preserves the symmetries
of the three-dimensional space
\be
ds^2 = dr^2 + e^{2 A(r)} \eta_{\m\n} dx^\m dx^\n \ ,\qq \m,\n=1,2,3\ .
\label{metriki}
\ee
As a result, the domain walls of the theory that preserve 1/2-supersymmetry,
correspond to solutions of the non-linear first-order system
of differential equations\footnote
{For a detailed explanation of how this system of first order
equations is derived 
from a general theory of scalars coupled to gravity and a potential 
obtained from a superpotential as in \eqn{dina1}, see section 3.1 of
\cite{BBDS}. We also note that, solutions to this system of differential 
equations, also solve the second order equations corresponding to the 
Lagrangian \eqn{laagra}.}
\newpage
\ba 
&& \sqrt{2} {d \phi_1 \over dr}  =
-{1\over 2} e^{\phi_1+\phi_+} + 2 e^{\phi_1} \sinh\phi_+ H_1^+ H_1^- 
+e^{-\phi_1} \left(e^{\phi_-} H_2^+ H_2^- + e^{-\phi_-} H_3^+ H_3^- \right) 
\, ,
\nonumber\\
&&\sqrt{2} {d \phi_2 \over dr}   = 
-{1\over 2} e^{\phi_1+\phi_+} + 2 e^{\phi_1} \cosh\phi_+  H_1^+ H_1^-
-e^{-\phi_1} \left(e^{\phi_-} H_2^+ H_2^- - e^{-\phi_-} H_3^+ H_3^- \right) 
\, ,
\nonumber\\
&& \sqrt{2} {d \phi_3 \over dr}   =  
-{1\over 2} e^{\phi_1+\phi_+} + 2 e^{\phi_1} \cosh\phi_+  H_1^+ H_1^-  
+e^{-\phi_1} \left(e^{\phi_-} H_2^+ H_2^- - e^{-\phi_-} H_3^+ H_3^- \right) 
\, ,
\nonumber\\
&&\sqrt{2} {d H_1^\pm\over dr}   = 
e^{\phi_1}\sinh\phi_+\, \left( 1+ 4(H_1^\pm)^2\right) H_1^\mp
\nonumber\\
&& \phantom{xxxxxxxxxxx}
-2e^{-\phi_1} 
\left(e^{\phi_-} H_2^+ H_2^- + e^{-\phi_-} H_3^+ H_3^- \right) H^\pm_1
\, ,
\label{hh1}\\
&&\sqrt{2} {d H_2^\pm\over dr}    = 
- {1\ov 2 }  e^{-\phi_1+\phi_-} H_2^\mp 
+4 e^{\phi_1}\sinh\phi_+\,  H_1^+ H_1^- H_2^\pm 
\nonumber\\
&&\phantom{xxxxxxxxxxx}
-2e^{-\phi_1}\left(e^{\phi_-} H_2^+ H_2^- 
+ e^{-\phi_-}H_3^+ H_3^-\right)H_2^\pm
\, ,
\nonumber\\
&& \sqrt{2} {d  H_3^\pm\over dr}   = 
-{1\over 2} e^{-\phi_1-\phi_-} H_3^\mp 
+4 e^{\phi_1}\sinh\phi_+\,  H_1^+ H_1^- H_3^\pm  
\nonumber\\
&& \phantom{xxxxxxxxxxx}
-2e^{-\phi_1} \left(e^{\phi_-} H_2^+ H_2^- 
+ e^{-\phi_-} H_3^+ H_3^- \right) H_3^\pm 
\, ,
\nonumber
\ea
where we found it more convenient to work with the fields 
$H_A^\pm$ instead of $z_A^\pm$
and we defined $\phi_\pm = \phi_2 \pm \phi_3$. 
As soon as a solution has been found, the conformal factor of the 
metric can be obtained by a simple integration of the resulting 
expression for the superpotential $W$, since
\be
{dA\ov dr} = {1\ov \sqrt{2}} W\ .
\label{hew1}
\ee
This system of equations for 9 real fields is rather
complicated and difficult to solve in general.
Instead, we will focus attention on subsectors obtained by
consistent truncations of the field content. 
Each one of these consistent truncations  
results into a system of 4 first-order equations for an equal number of
real fields that  
correspond to the various type-II and heterotic theories. 
It will 
turn out that the general solutions 
for all type-II theories can be found explicitly (section 2). 
For the heterotic case (section 3)
a general solution cannot be given in closed form.
However, it is possible to extract 
the general behaviour of the fields in the strong and weak coupling regions 
as well as around certain critical points.
An explicit solution is given in the limit of infinitely large left-winding 
field.
The physical features of our solutions are explained in the appropriate
places in the text.
In section 4 we summarize our results, we present some directions for
future work and also generally comment on various issues 
related to our approach to superstrings at finite temperature.

\section{Type II string theories}
\setcounter{equation}{0}

The simplest truncations of
the domain wall equations lead to different type-II sectors. 
We first examine the type-IIA and type-IIB theories, which can be
treated simultaneously,
and then examine a hybrid type-II sector, which describes type-II strings
at the self-dual radius. Both cases turn out to be exactly solvable.

\subsection{Type-IIA and IIB sector}
The type-IIA and type-IIB sectors of the theory are obtained by setting
\ba
&& z_1^\pm =  z_2^\pm = 0 ~, ~~~~~ {\rm for ~ type\ IIB}\ , \\
&& z_1^\pm =  z_3^\pm = 0 ~, ~~~~~ {\rm for ~ type\ IIA} \ ,
\ea
in which case $H_1^\pm = H_2^\pm = 0$ and $H_1^\pm = H_3^\pm = 0$,
respectively. It follows 
from \eqn{hh1} that $\phi_1$ equals $\phi_2$ 
or $\phi_3$, respectively,
up to an irrelevant additive constant which we will ignore.
Then, it is convenient to set 
\be
\phi_1 = \phi_2 ={\phi\ov 2} ~, \qq \phi_3 = \chi \ ,\qq
z_3^\pm = {\rm tanh}\left({\omega_\pm \over 2}\right) \ ,\qq
 {\rm for ~ IIB}  \ ,
\ee
or 
\be
\phi_1 = \phi_3 = {\phi\ov 2} \ ,\qq \phi_2 = \chi \ ,\qq
z_2^\pm = {\rm tanh}\left({\omega_\pm \over 2}\right) \ ,\qq {\rm for ~ IIA}\ ,
\ee
where we have introduced in either case the fields $\omega^\pm$.
Then, $2 H^\pm_2= \sinh \om_\pm$ for the type-IIA theory and similarly
$2 H^\pm_3= \sinh \om_\pm$ for the type-IIB theory. 
The temperature field in type-II units 
is $T\sim e^{ \phi}$ and the string 
coupling is $g_s\sim e^\chi$.\footnote{For explanations on the
identifications of the various fields as temperature, string coupling and 
windings, see \cite{adk,BBDS}.}
As in \cite{BBDS} 
we can treat both cases together because the superpotential and the
potential assume the same form for the type-IIA and type-IIB theories. 
Hence, no distinction will be made in the following between the
type-IIA or type-IIB theories.
The kinetic terms in the Lagrangian 
for the fields $\chi$, $\phi$ and $\om$ 
assume the form 
\be
{\cal L}_{\rm kin}= -\ha (\del \chi)^2 -{1\ov 4} (\del\phi)^2 -{1\ov 4} 
(\del \om_+)^2 -{1\ov 4} (\del \om_-)^2\ .
\label{llkin}
\ee
In that respect the definition for the field $\phi$ in this paper and in 
\cite{BBDS} differ by a factor of $\sqrt{2}$.

Explicit calculation shows that in terms of the new variables the
superpotential \eqn{supereal} becomes
\be
W_{{\rm II}} = {1 \over 2} e^{\chi} \left(e^{ \phi} + 
{1 \over 2} e^{-\phi} {\rm sinh} \om_+ \sinh\om_-  \right) \ , 
\label{hsj3}
\ee
whereas the corresponding potential \eqn{dina1} takes the form
\be
V_{{\rm II}} = {1 \over 64} e^{2 \chi} \left(e^{-2 \phi}
(\cosh 2\om_+ \cosh 2\om_- -1) -16 \sinh\om_+
\sinh \om_- \right)\ .
\label{hsj}
\ee
Using the truncated superpotential $W_{II}$,
the type-II domain walls obey the system of first-order
equations 
\ba
{d \chi \over dr} & = & -{1 \over 2 \sqrt{2}} e^{\chi} 
\left(e^{ \phi} + 
{1 \over 2} e^{-\phi} {\rm sinh} \om_+ \sinh\om_-\right)\ ,
\nonumber\\
{d \phi \over dr} & = & -{1 \over \sqrt{2}} e^{\chi} \left(e^{\phi} 
- {1 \over 2} e^{-\phi} \sinh \omega_+ \sinh \om_-
 \right) , 
\label{syII}\\
{d \omega_\pm \over dr} & = & -{1 \over 2 \sqrt{2}} 
e^{\chi-\phi}\sinh \om_\mp \cosh \om_\pm ~.
\nonumber
\ea
Since this system of equations is
invariant under $\om_\pm\to -\om_\pm$ we will consider the case 
$\om_+\ge 0$ without any loss of generality. 
Let us also mention that, for 
$\om_+=\om_-$ all the expressions so far in this section go over to
the corresponding expressions in section 4.1 of \cite{BBDS}.

\subsubsection{The general solution} 
The system of eqs. \eqn{syII} can be
completely integrated. 
First, we easily can show that the two winding fields are related as
\ba
&& \om_+ = \om\ ,
\nonumber\\
&& \cosh \om_- = \l \cosh\om \ , \qq \sinh\om_- =\pm \sqrt{\l^2 \cosh^2\om
-1}\ ,
\label{lala}
\ea
where $\l$ is a positive constant and where, for notational convenience
we have introduced the field $\om$. 
This relation completely determines $\om_-$ in terms of $\om_+$ up to 
an important sign. In fact, the different cases 
corresponding to the plus and minus signs are associated with
different categories of solutions.
In particular, when $\om_+\om_-<0$ the potential \eqn{hsj} is manifestly 
positive and there are cannot be any solutions, supersymmetric or not,
that lead to tachyonic instabilities.
We may restrict to $\l\geq 1$ since for $\l<1$ the 
r\^oles of $\om_+$ and $\om_-$ are interchanged.

It is convenient to present the solution for the rest of the fields and the
metric by treating the field $\om$ as an independent variable,
which is legitimate since the third equation in \eqn{syII} implies that 
$\om_+=\om$ is a monotonous function of $r$.
Then, the differential equation for $\phi$ can be easily integrated
and also the equation for $\chi$.
The resulting family of BPS solutions has
\be
e^{-2 \phi}  =  \pm \cosh^2\om\ \L(\om)\ ,\qq 
e^{2 \chi}  =   \cosh^2\om\ e^\phi\ ,
\label{chiII}\\
\ee
where the $+$ and $-$ signs correspond to the sign of the product $\om_+\om_-$.
The function $\L(\om)$ is given by 
\ba
\L(\om) &  = & C \ + \
2 \sqrt{y^2-(1+\l^2) y +\l^2} 
\ + \ (\l^2+1) 
\nonumber\\
&& \times \ \ln\left(\l^2+1 -2 y - 2 \sqrt{y^2-(1+\l^2) y +\l^2} \ov \l^2-1
\right) \ ,\qq y=1/\cosh^2 \om \ ,
\nonumber
\ea
and is parametrized by an arbitrary integration constant $C$.
Another multiplicative integration constant on the right hand side of the 
expression for $e^{2\chi}$ in \eqn{chiII} has been omitted,
since it can always be absorbed into trivial field redefinitions.
The physical interpretation of the solutions  
depends crucially on a critical value for the constant $C$.
Before we examine this issue let us mention that 
the equation for the conformal factor of the metric \eqn{metriki} is easily 
integrated and gives $A=-\chi$ (up to a constant that can be absorbed into
a redefinition of the $x^\m$'s). Hence, the
metric takes the form
\be
ds^2 ={8 e^{ \phi}\ov (\l^2\cosh^2\om -1) \cosh^4\om}\ d\om^2 
\ +\ {e^{-\phi}\ov \l \cosh^2\om}\ \eta_{\m\n} dx^\m dx^\n \ .
\label{metII}
\ee
The relation between the variables $r$ and $\om$ in \eqn{metriki}
and \eqn{metII} is given by the relation of 
differentials in the last eq. in \eqn{syII}.
This integration can be performed to yield $r(\om)$ in terms of special 
functions, but it cannot be inverted to get the explicit dependence 
of $\om(r)$, except in 
some limiting cases, as we will see.

In order to find the critical value for the integration constant $C$ let us 
first note the limiting values for the function $\L(\om)$, namely
\ba
&& \L(0)= C\ , \qq \L(\infty) = C -C_{\rm crit}\ ,
\nonumber\\
&& C_{\rm crit} =  
(\l^2+1) \ln\left(\l+1\ov \l-1\right)- 2\l\ >\ 0 \ ,\qq \forall\
\l \geq 1\ .
\label{limm}
\ea
It is easily seen that in the interval $[0,\infty)$ the function $\L(\om)$
is monotonously decreasing between the above two values. 

The different cases, as we have seen, correspond to the product $\om_+\om_-$ 
being positive or negative. If any of the winding fields 
$\om_\pm$ is taken zero, that implies that the other one is zero as well.
The resulting solution in that case was presented in \cite{BBDS} and will 
not be repeated here.

\medskip
\no
\underline{$\om_+ \om_- >0$:} Then,
the reality condition for the various fields 
requires that $C>0$. For $\om\to 0$ the fields $\phi$, $\chi$ 
and $\om_-$ reach some constant values determined by $C$ and the metric becomes
the four-dimensional Euclidean flat metric.\footnote{This is a 
universal behaviour and it should be contrasted with 
the situation in \cite{BBDS} where there was no flat limit when $\om\to 0$.
The reason, as explained below, is that the two solutions are related 
by a singular limit.} 
However, whether $\L(\infty)$ is positive, negative or zero
distinguishes three different cases: If $C>C_{\rm crit}$ 
the winding field $\om$ takes 
values in the entire real line, i.e. $\om\in [0,\infty)$.
In this interval, the temperature field $T\sim e^\phi$ first increases
until a maximum value and then it decreases to zero as $\om\to \infty$.
In contrast, the string coupling $g_s\sim e^\chi$ is a monotonously 
increasing function of $\om$ in the entire interval.
If $C=C_{\rm crit}$, then still $\om\in [0,\infty)$, but now 
both the temperature and the string coupling are monotonously 
increasing functions of $\om$ from a constant at $\om=0$, to infinity 
as $\om\to \infty$.
If $C<C_{\rm crit}$, then there exists a maximum value for the winding
field, $\om_{\rm max}$, beyond which the fields become imaginary.
As a function of $\om\in [0,\om_{\rm max})$ both the temperature $T$ and the 
string coupling $g_s$ are monotonously increasing from their
constant values at $\om=0$ to infinity as $\om\to \om_{\rm max}^-$.

It is important to work out the expressions for the various fields and 
the metric near the maximum value of $\om$, whether that is finite of infinite.
In terms of the variable $r$
\ba
&&C>C_{\rm crit}: \qq e^{-\om}\sim r^{2/7}\ ,\qq e^{\phi}\sim r^{2/7}\ ,\qq
e^{2\chi} \sim r^{-2/7}\ ,
\nonumber\\
&&C=C_{\rm crit}: \qq e^{-\om}\sim r^{2/5}\ ,\qq e^{\phi}\sim r^{-2/5}\ ,\qq
 e^{2\chi} \sim r^{-6/5}\ ,
\label{hgh3}\\
&&C<C_{\rm crit}: \qq \om_{\rm max}-\om \sim r^{4/3}\ ,
\qq e^{\phi}\sim e^{2\chi}
\sim r^{-2/3}\ ,
\nonumber
\ea
as $r\to 0^+$.
The metric takes the form 
\be
ds^2\simeq dr^2 +(\const)\ r^{2\n}\eta_{\m\n} dx^\m dx^\n\ ,\qq
{\rm as} \quad r\to 0^+\ ,
\label{hghg3}
\ee
where we have introduced the constant parameter $\n=1/7$, $3/5$ and
$1/3$ corresponding to $C$ being larger, equal or smaller 
than $C_{\rm crit}$, respectively.
The value at $r=0$ corresponds to a naked 
curvature singularity. In order to fully understand the nature of this
singularity ultimately we have 
to go beyond the effective supergravity description. Since, presently this
is not feasible, we can study instead whether the propagation of test quantum
particles in our backgrounds is well defined. This requires a unique time
evolution of initial data or equivalently a unique self-adjoint extension
of the wave-operator at the singularity \cite{wald,homa,ishibashi}.
This criterion has been successfully applied within the AdS/CFT correspondence
in many backgrounds related to the Coulomb branch of supersymmetric 
gauge theories in the sense that the results are in agreement with field 
theory expectations \cite{BrSf}. This is essentially the reason that we 
trust this criterion in our case, where there is no independent information
on the physical status of our solutions.
For more discussion on these issues and an adaptation to metrics with
Euclidean signature and in particular of the form \eqn{hghg3},
as in our case, see section 6 of \cite{BBDS}. 
It turns out that this criterion 
requires that for the case of backgrounds that take place in a infinite
interval of $r$, i.e., $r\in [0,\infty)$, the constant $\n\ge 1/3$.
In the cases where the solutions takes place at a finite interval of $r$
with two singularities, one at $r=0$ and another one at $r=r_0$,
the criterion requires that $\nu\ge 1/3$ at one singularity and 
$\nu < 1/3$ at the other. 
Returning to our solutions we see that only the 
family of solutions with $C\leq C_{\rm crit}$ is `physical'.

\medskip
\no
\underline{$\om_+ \om_- <0$:} Then, if $C<0$, reality of 
the fields allows the full range for $\om\in [0,\infty)$. It turns out that 
$e^\phi$ ($e^{2 \chi}$) decreases (increases) monotonously from a constant 
value to zero (infinity). 
The behaviour of the various fields 
when $\om\to \infty$ is 
\be
e^{-\om}\sim e^\phi \sim e^{-2\chi}\sim (r_0-r)^{2/7}\ ,\qq {\rm as} \quad
r\to r_0^-\ ,
\label{42g}
\ee
where $r_0$ is a positive constant.
For the metric we have the form \eqn{hghg3} with $\n=1/7$ and $r$
replaced by $r_0-r$.

If $0<C< C_{\rm crit}$, there exists a minimum
value $\om_{\rm min}$ below which the fields become imaginary.
In the interval $[\om_{\rm min},\infty)$, $e^\phi$ is a monotonously
decreasing function from infinity to zero. Instead, $e^{2 \chi}$ between 
becoming infinity at the two interval-ends, reaches a minimum value.
For $\om\to \om_{\rm min}^+$ the various fields behave as 
\be
\om - \om_{\rm min} \sim r^{4/3}\ ,\qq \e^\phi \sim e^{2\chi}\sim r^{-2/3}\ ,
\qq {\rm as} \quad r\to 0^+\ ,
\label{24g}
\ee
whereas the metric behaves as in \eqn{hghg3} with $\n=1/3$. 
For $\om\to \infty$ the behaviour of the fields and the metric is as in the 
case with $C<0$ above.
We see that according to our criterion  
only the family of solutions with $0<C< C_{\rm crit}$ is `physical'.

Finally, for $C\geq C_{\rm crit}$ it is impossible to satisfy the reality 
condition for the fields.

At this point let us note how a disturbing, from a physical view point, 
feature of our solutions for $0<C<C_{\rm crit}$ 
is resolved to our advantage. 
For either sign of $\om_+ \om_-$ the winding field $\om$ is not allowed to take
arbitrary positive values, since there is either an upper or a lower bound.
As in General Relativity we may try to continue the solution
across the singularity occurring at this boundary value of $\om$.
This makes sense also 
from the point of view that the singularity has been characterized 
as `physical' according to our criterion. 
To be concrete, consider the case with $\om_+ \om_->0$ and analytically 
continue 
from $\om<\om_{\rm max}$ to $\om>\om_{\rm max}$. 
Such an analytic continuation 
brings us to the solution with $\om_+ \om_- <0$ with the identification of 
the two constants $\om_{\rm max}$ and $\om_{\rm min}$. 
This is an elementary 
mechanism of changing the sign of the 
winding number in our effective supergravity 
action approach.\footnote{Interestingly, 
the flipping of the sign of $\om_+ \om_-$
corresponds to the complexification of the fields as $e^\phi\to i e^\phi$
and $e^\chi \to -i e^\chi$, in such a way that the system of eqs. \eqn{syII}
and the solution \eqn{chiII} remain real.} 

\subsubsection{The truncation $\om_- =\pm \om_+$}

Two cases of particular interest arise when the constant $\l=1$.
From \eqn{lala} we have that, either $\om_-=\om_+=\om$, or
$\om_-=-\om_+=-\om$. In these cases the solution reads
\be
e^{-2 \phi} =\pm 2\left( \cosh^2\om (\ln \coth^2\om + c) -1\right)\ ,
\label{kj2}
\ee
where the plus (minus) sign corresponds to equal (opposite) winding fields.
This solution can be found by solving the system of eqs. \eqn{syII}
after we set $\om_-=\pm \om_+=\pm \om$, as this is a consistent truncation of
the system of equations. It can also be found by taking the limit $\l\to 1^+$ 
in the solution \eqn{chiII}, after we regularize the resulting infinite 
expression by also shifting the integration constant $C$ by an infinite
positive constant as
$C=2 (c-1) -2 \ln\left(\l-1\ov 2\right)$. Since, in this
limit $C>0$ and $C-C_{\rm crit}=2 c$, the solution for 
$\om_- =\om_+$ should be characterized by whether  
the constant $c$ is positive, zero or negative. Instead, for $\om_- =-\om_+$
only $c<0$ should be allowed.
Indeed, the solution for the case $\om_+=\om_-$ was found in \cite{BBDS} 
with
behaviour that depended crucially on whether the constant $c$ 
was positive, zero or negative in accordance with our general
discussion.
We will not elaborate in this case, except for the following remark: The 
solution in \cite{BBDS},
with $\om_-=\om_+$, does not have as a limiting case for $\om\to 0$
flat Euclidean space with all fields constant. On the other hand we see that 
this was the limiting behaviour of the general solution \eqn{chiII}.
Then, a paradox seems to arise since \eqn{kj2} is a limit of \eqn{chiII}.
The resolution to that comes with the realization,
after a closer inspection to the 
various expressions involved and using
the fact that the constant $C$ is shifted 
by an infinite amount, that the two limits $\om\to 0$ and $\l\to 1^+$ 
do not commute, since the latter limit is singular.

In the case with $\om_+=-\om_-$ the 
reality conditions on the fields imposes that $c<0$ and that there is a
minimum value for the winding field, i.e., 
$\om \geq \om_{\rm max}$, in accordance to our general discussion before.
In the limit of small or large $|c|$ the
constant $\om_{\rm min}$ can be found analytically
\be
\om_{\rm min} = \left\{\begin{array}{lll}
-\frac{1}{4} \ln\left(-c\ov 8\right) & \ \ {\rm for}\  & c \to 0^-\ ,
 \\
e^{c-1\ov 2} & \ \ {\rm for}\  & c \to - \infty  \ ,
\end{array} 
\right.
\label{lima6}
\ee
whereas for intermediate values of $c$, $\om_{\rm max}$ 
ranges between the above two extremes.
In this range, $e^{2 \phi}$ is a decreasing function of $\om$.
Instead, $e^{2\chi}$ can acquire a minimum value at some 
$\om=\om_0>\om_{\rm min}$ 
which, as before, in the limit of large $|c|$,
can be computed analytically
\be
\om_{\rm 0} = \left\{\begin{array}{lll}
-\frac{1}{4} \ln\left(-c\ov 12\right) & \ \ {\rm for}\  & c \to 0^-\ ,
 \\
2^{-1/4} (-c)^{-3/4} & \ \ {\rm for}\  & c \to - \infty  \ ,
\end{array} 
\right.
\label{lim3}
\ee
whereas for intermediate values of $c$, $\om_0$ 
ranges between the two above extremes.
The corresponding value for the field $\phi$ is
given by
\be
e^{\phi}\Big|_{\om=\om_0}={1\ov \sqrt{2}} \sinh\om_0\ .
\ee

The asymptotic behaviour of the various fields 
near $\om=\om_{\rm min}$
are given by \eqn{24g}, whereas for the metric by \eqn{hghg3} with $\n=1/3$.
Similarly, for $\om\to \infty$ the behaviour of the fields is given  by 
\eqn{42g} and for the metric by \eqn{hghg3} with $\n=1/7$, after replacing
$r$ by $r_0-r$.

\subsubsection{The truncation in the limit $\om_- \to \pm \infty$}

In is interesting to have a closer look to the case where one of the winding
fields becomes much larger than the other one. 
Without loss of generality we choose $\om_-\gg \om_+$.
In this case consider the rescaling followed by the limit
\be
\phi \to \phi +\ha \ln(\l/4)\ , \qq \chi \to \chi -\ha \ln(\l/4)\ , \qq
\l\to \infty\ ,
\ee
and also substitute $\sinh\om_-\simeq \l \cosh\om$ using \eqn{lala}.
Then the system of equations \eqn{syII} becomes 
\ba
{d \chi \over dr} & = & -{1 \over 2 \sqrt{2}} e^{\chi} 
\left(e^{ \phi} \pm e^{-\phi} \sinh 2 \om\right)\ ,
\nonumber\\
{d \phi \over dr} & = & -{1\ov  \sqrt{2}} e^{\chi} \left(e^{\phi} 
\mp  e^{-\phi} \sinh 2 \omega  \right)\ , 
\label{sy23I}\\
{d \omega \over dr} & = & \mp \sqrt{2} 
e^{\chi-\phi} \cosh^2\om\ .
\nonumber
\ea
We emphasize that in this limit the potential \eqn{hsj} and the superpotential
\eqn{hsj3} are well defined as one can easily check. In addition, the 
kinetic term for the field $\om$ in \eqn{llkin} is also well defined and
becomes $-(1+\tanh^2\om)(\del\om)^2/4$. Hence, the limit $\om_-\to \pm \infty$
can be considered as a consistent truncation of the theory, much like the 
truncation with $\om_-=\pm \om_+$.

The solution of \eqn{sy23I} in parametric form is
\ba
e^{-2\phi} & = & 
\pm \cosh^2\om \left(a-{1\ov 3}\tanh\om (3-\tanh^2\om)\right)\ ,
\nonumber\\
e^{2\chi} & = & \cosh^2 \om\ e^\phi\ .
\label{hge5}
\ea
This solution can be also obtained from \eqn{chiII} in the limit 
$\l\to \infty$, after choosing the 
constant equal to $C=4 a/\l $. 
The different behaviours are found by 
noticing also that, in this limit $C-C_{\rm crit} =4(a-2/3)/\l$.
Then, for $\om_-\to +\infty$, we get different behaviours depending 
on whether $a>2/3$, $a=2/3$ or $0< a< 2/3$. 
This resembles the behaviour of the 
general solution for $\om_+\om_->0$. 
For $\om_-\to -\infty$, different behaviours arise depending on 
whether $a<0$ or $0< a< 2/3$ and resemble the behaviour of the general 
solution for $\om_+\om_-<0$.

\subsection{A hybrid type-II sector}
There is another truncation of the domain wall equations with  
$H_1^\pm=0$, as in type-II, which is exactly solvable. 
We set for this purpose 
$H_2^\pm = \pm H_3^\pm$ and observe that the full system of equations is
consistent provided that $\phi_2 = \phi_3$. As in \cite{BBDS}, we call
this sector a hybrid of type-IIB and type-IIA and amounts to choosing 
the self-dual radius so that there
is no distinction between the type-IIA and type-IIB theories.

We proceed further by setting
\be
\phi_1 = \chi ~,\qq \phi_2 = \phi_3 = {1 \over 2}\phi ~,\qq
z_2^\pm = \pm z_3^\pm = {1 \over \sqrt{2}} {\rm tanh}{\omega^\pm \over 2}
\ee
and so $H_2^\pm = \pm H_3^\pm = {1 \over 2\sqrt{2}} {\rm sinh}\omega_\pm$.
The kinetic terms of the fields $\chi$, $\phi$ and 
$\omega_\pm$ assumes the same form as in \eqn{llkin}.
As before, the temperature field is $T\sim e^{\phi}$ and the string coupling 
is $g_s \sim e^\chi$.
Then, the superpotential \eqn{supereal} and the corresponding potential become 
\be
W_{{\rm hyb}} = {1 \over 2}\left(e^{\chi + \phi} 
+{1 \over 2} e^{-\chi}  {\rm sinh}\om_+\sinh\om_- \right)
\ee
and 
\be
V_{{\rm hyb}} = {1\ov 128} e^{-2 \chi} (\cosh 2\om_+ \cosh 2\om_- + 
\cosh\om_+ + \cosh\om_- -3)-{1\ov 4} e^\phi \sinh\om_+ \sinh\om_-\ ,
\ee
respectively.

The truncated system of equations takes a simpler form in this case, namely 
\ba
{d \chi \over dr} & = & -{1 \over 2 \sqrt{2}} 
\left(e^{\chi + \phi} 
-{1 \over 2} e^{-\chi}  {\rm sinh}\om_+\sinh\om_- \right)\ ,
\nonumber\\
{d \phi \over dr} & = & -{1 \over \sqrt{2}} e^{\chi + \phi} \ ,
\label{syhybII}\\ 
{d \omega_\pm \over dr} & = & -{1 \over 2\sqrt{2}}e^{-\chi} \sinh \om_\mp
\cosh\om_\pm  ~ .
\nonumber 
\ea
The equation for the 
conformal factor of the metric \eqn{metriki} can be easily integrated; 
it gives 
$A=-\chi-\ha \ln (\cosh \om_+ \cosh \om_-)$, up to a constant that can 
be absorbed into a redefinition of the $x^\m$'s. Notice that 
the result is not the same as in the 
genuine type-II case that we examined before.
The system of equations \eqn{syhybII} is 
invariant under $\om_\pm\to -\om_\pm$, and as before we only need to
consider the case of
$\om_+\ge 0$. 
Similarly to the genuine type-II case, 
for $\om_+=\om_-$ all the expressions so far in this section go over to
the corresponding expressions in section 4.2 of \cite{BBDS}.

\medskip
\no
\underline{General solution}: As before, 
it is easily seen that the winding fields $\om_\pm$ are
related by \eqn{lala} and similarly we take the constant $\l\geq 1$ 
with no loss of generality.
The general solution of the remaining equations gives 
the following family of BPS solutions with  
\ba
e^{-2\phi} & = & \pm 4 \sqrt{\l} \Big( F(\varphi,1/\l) -
E(\varphi,1/\l) + E(1/\l) - K(1/\l) + B  \Big)\ ,
\nonumber\\
e^{2\chi} & = & {e^\phi\ov \l^{1/2} 
\cosh\om }\ ,\qq \sin\varphi = 1/\cosh\om\ ,
\label{1jh2}
\ea
where $B$ is a constant of integration and where we have used the standard 
notation for the complete and incomplete elliptic integrals of the first and
the second kind. As before, the $+$ and $-$ signs correspond to the sign
of $\om_+\om_-$. Also we found that the metric takes the form
\be
ds^2 = {8\l^{-1/2} e^{\phi}\ov \sqrt{\l^2\cosh^2\om -1} \cosh^3\om}\ d\om^2\ +
\ { e^{-\phi}\ov \l^{1/2} \cosh\om}\ \eta_{\m\n} dx^\m dx^\n \ .
\label{metHyb}
\ee
The connection between the variables $r$ and $\om$ in \eqn{metriki}
and \eqn{metHyb} is given by a relation of the corresponding differentials.
As before, the integration can be performed and yields $r(\om)$ in a
closed form,
but it cannot be inverted to get $\om(r)$, apart from
a few limiting cases, as we will see next.

To investigate the properties of the solution we note 
the limiting values
\ba
&& e^{-2\phi}|_{\om=0} = \pm 4 B \ , 
\qq e^{-2\phi}|_{\om=\infty} = \pm 4(B-B_{\rm crit})
\ ,
\nonumber\\
&& B_{\rm crit} = K(1/\l) - E(1/\l)\ >\ 0\ ,\qq \forall\ \l\geq 1 \ .
\label{gfde}
\ea

As before, the sign of $\om_+\om_-$ distinguishes different types of 
solutions.

\medskip
\no
\underline{$\om_+ \om_- >0$:} Then,
the reality condition for the various fields 
requires that $B>0$. In the limit $\om\to 0$ the fields $\phi$, $\chi$ 
and $\om_-$ approach some constant values determined by $B$ and 
the metric becomes
the four-dimensional Euclidean flat metric (see also footnote 4).
However, whether $B$ is larger, equal or less than $B_{\rm crit}$ 
distinguishes three different cases: If $B>B_{\rm crit}$ then 
the winding field $\om$ can take values in the entire real line, 
i.e. $\om\in [0,\infty)$.
In this interval, the temperature field $T\sim e^\phi$ increases
monotonously between the two constant values that can be read off from
\eqn{gfde}.
The string coupling $g_s\sim e^\chi$ goes from a constant value to zero
after reaching a maximum.
If $B=B_{\rm crit}$, then still $\om\in [0,\infty)$, but now 
both the temperature and the string coupling are monotonously 
increasing functions of $\om$ and go from a constant to infinity.
If $B<B_{\rm crit}$, then there exists a maximum value for the winding
field, $\om_{\rm max}$, beyond which the fields are not real.
As a function of $\om\in [0,\om_{\rm max})$ both the temperature $T$ and the 
string coupling $g_s$ are monotonously increasing from their constant values 
at $\om=0$, to infinity as $\om\to \infty$.

The behaviour of the various fields and 
the metric near the maximum value of $\om$ (finite of infinite), 
in terms of the variable $r$, is given by
\ba
&&B>B_{\rm crit}: \qq e^{-\om}\sim e^{2\chi}\sim r^{2/5}\ ,\qq e^{\phi}\simeq
\ha (B-B_{\rm crit})^{-1/2} + (\const)\ r^{6/5}\ ,
\nonumber\\
&&B=B_{\rm crit}: \qq e^{-\om}\sim r^{4/7}\ ,\qq e^{\phi}\sim r^{-6/7}\ ,\qq
 e^{2\chi} \sim r^{-2/7}\ ,
\label{hgh31}\\
&&B<B_{\rm crit}: \qq \om_{\rm max} -\om \sim r^{4/3}\ ,
\qq e^{\phi}\sim e^{2\chi}
\sim r^{-2/3}\ ,
\nonumber
\ea
as $r\to 0^+$. The metric takes the form \eqn{hghg3} with $\n=1/5$, $5/7$ and
$1/3$ corresponding to $B$ being larger, equal or smaller 
than $B_{\rm crit}$, respectively.
As before, $r=0$ corresponds to a naked 
curvature singularity. 
According to our criterion only the  
family of solutions with $B\leq B_{\rm crit}$ is `physical'.

\medskip
\no
\underline{$\om_+ \om_- <0$:} Then, if $B<0$ the reality condition on
the fields is obeyed for $\om\in [0,\infty)$. It turns out that 
$e^\phi$ and $e^{2 \chi}$ decrease monotonously from a constant 
value to a constant and a zero value, respectively. 
The behaviour of the various fields 
when $\om\to \infty$ is 
\be
e^{-\om}\sim e^{2\chi}\sim  (r_0-r)^{2/5}\ ,
\qq e^\phi \simeq \ha (B_{\rm crit}-B)^{-1/2}
+(\const) \ (r_0-r)^{6/5}\ ,
\label{42g1}
\ee
where $r_0$ is a positive constant. 
The metric we have the form \eqn{hghg3} with $\n=1/5$ and $r$
replaced by $r_0-r$.

If $0<B< B_{\rm crit}$, there exists a minimum
value $\om_{\rm min}$ below which the fields are not real.
In the interval $[\om_{\rm min},\infty)$, $e^\phi$ and $e^{2 \chi}$
are monotonously
decreasing functions from infinity to a constant and a zero value, 
respectively. 
For $\om\to \om_{\rm min}^+$ the various fields behave as 
\be
\om - \om_{\rm min} \sim r^{4/3}\ ,\qq \e^\phi \sim e^{2\chi}\sim r^{-2/3}\ ,
\label{24g1}
\ee
whereas the metric behaves as in \eqn{hghg3} with $\n=1/3$. 
For $\om\to \infty$ the behaviour of the fields and the metric is as in the 
case with $B<0$ above.
We see that, according to our criterion  
only the family of solutions with $0<B< B_{\rm crit}$ is `physical'.

For $B\geq B_{\rm crit}$ it is impossible to satisfy the reality 
condition for the fields.

Let us mention also that the change of the sign of the 
winding number mechanism,
when we analytically continue through the singularity, that 
was exposed for the genuine type-II case in the paragraph
just before section 2.1.2, applies also in this case for $0<B<B_{\rm crit}$.

\subsubsection{The truncation $\om_+=\pm \om_-$}

As for the genuine type-II theories, the case 
when the constant $\l=1$ is of particular interest. Then,
from \eqn{lala} we have that either $\om_-=\om_+=\om$, or
$\om_-=-\om_+=-\om$. In these cases the solution reads
\be
e^{-2\phi}  =  \pm \left(b + 
4 \left({\rm \ln}\left({\rm coth}{\omega \over 2}
\right) - {1 \over {\rm cosh} \omega}\right) \right)\ ,
\label{g87}
\ee
where $b$ is an integration constant.
Being a consistent truncation, this solution can be found 
by solving the system of eqs. \eqn{syhybII}
after we set $\om_-=\pm \om_+=\om$. It can also be found by taking the 
limit $\l\to 1^+$ in the solution \eqn{1jh2}, after we 
regularize the resulting infinite 
expression by also shifting the integration constant $B$ by an infinite
positive constant as
$B=b/4 -1 -\ha \ln\left(\l-1\ov 8\right)$. Since in this limit $B>0$ and
$B-B_{\rm crit}=b/4$ the solution for $\om_-=\om_+$ should be characterized
of whether $b$ is positive, negative or zero, whereas for $\om_-=-\om_+$
only $b<0$ should be allowed.
The solution for the case $\om_+=\om_-$ was found in \cite{BBDS} and the
behaviour of the solution depended crucially on whether the constant $b$ 
was positive, zero or negative in accordance
with our expectations. Since the limiting procedure that leads to \eqn{g87}
from the general solution \eqn{1jh2} is singular,
their corresponding behaviours as $\om\to 0$ are different.

Turning to the case with $\om_-=-\om_+=-\om$, we see that 
the reality condition for the fields $\chi$ and $\phi$ requires that $b<0$
again in accordance with our expectations and a minimum value
for the winding field, i.e., 
$\om \geq \om_{\rm min}$.
In this range, $e^{2\chi}$ and $e^{\phi}$ are monotonically 
decreasing functions of $\om$.
In the limit of small and large 
$|b|$ we find for $\om_{\rm min}$ the analytic 
expressions
\be
\om_{\rm min} = \left\{\begin{array}{lll}
-\frac{1}{3} \ln\left(-3b\ov 32\right) & \ \ {\rm for}\  & b \to 0^-\ ,
 \\
2 e^{b/4-1} & \ \ {\rm for}\  & b \to - \infty  \ .
\end{array} 
\right.
\label{lhyyba6}
\ee
For intermediate values of $b$, $\om_{\rm min}$ ranges between the 
above two extremes.

The asymptotic behaviour of the various fields 
near $\om=\om_{\rm min}$
is given by \eqn{24g1}, whereas for the metric by \eqn{hghg3} with $\n=1/3$.
Similarly, for $\om\to \infty$ the behaviour of the fields is given  by 
\eqn{42g1} and for the metric by \eqn{hghg3} with $\n=1/5$.

\subsubsection{The truncation in the limit $\om_-\to \pm \infty$}

Let us consider now the limit of large left-winding field $\om_-$
and in the system of eqs. \eqn{syhybII} perform the rescaling followed by the
limit
\be
\phi\to \phi +{3\ov 4} \ln\l\ , \qq \chi\to \chi +{1\ov 8}\ln\l\ ,
\qq r\to r \l^{-7/8}\ ,
\ee
and also substitute $\sinh\om_-\simeq \l \cosh\om$ using \eqn{lala}. 
We find the result
\ba
{d \chi \over dr} & = & -{1 \over 2 \sqrt{2}} 
\left(e^{\chi + \phi} 
\mp {1\ov 4} e^{-\chi}  \sinh 2\om \right)\ ,
\nonumber\\
{d \phi \over dr} & = & -{1 \over \sqrt{2}} e^{\chi + \phi} \ ,
\label{sy25II}\\ 
{d \omega \over dr} & = & \mp{1 \over 2\sqrt{2}}e^{-\chi}  \cosh^2\om  ~ .
\nonumber 
\ea
with solution 
\ba
e^{-2\phi} & = &\pm 2 \left( d  + \varphi - {\sinh\om \ov
 \cosh^2\om} \right)\ ,
\nonumber\\
e^{2\chi} & = & {e^\phi \ov \cosh \om}\ ,\qq  \sin \varphi = 1/\cosh\om\ .
\ea
As before, this limit can be considered as a consistent truncation since the
potential, the superspotential and the kinetic term for the field $\om$ 
are all well defined. Finally, we mention that this solution can be obtained 
from the general solution in the limit $\l\to \infty$ and after we set
$B=(d+\pi/2)/(2 \l^2)$. The different behaviours are found by 
noticing also that, in this limit $B-B_{\rm crit} = d/(2\l^2)$.
Then, for $\om_-\to +\infty$, we get different behaviours depending 
on whether $d>0$, $d=0$ or $-\pi/2< d< 0$. This resemble the behaviours of the 
general solution for $\om_+\om_->0$. 
For $\om_-\to -\infty$, different behaviours arise depending on 
$d<-\pi/2$ or $-\pi/2< d< 0$ and resemble the behaviour of the general 
solution for $\om_+\om_-<0$.

\section{Heterotic sector} 
\setcounter{equation}{0}
 
The heterotic limit is obtained by setting $H_2^\pm = H_3^\pm = 0$, while 
keeping $H_1^\pm$ free to vary. Then, introducing
new fields as 
\be
\phi_1 = \chi ~, ~~~~ \phi_2 = \phi_3 = {\phi \ov 2} ~, ~~~~ 
z_1^\pm = {\rm tanh} {\omega_\pm \over 2} ~, 
\ee
for which $2H_1^\pm = {\rm sinh} \omega_\pm$, 
we obtain the expression for the truncated superpotential 
\be
W_{{\rm het}} = {1 \over 2} e^{\chi} \left( e^{\phi} - 
{\rm sinh}\phi \sinh\omega_+ \sinh \om_- \right)\ , 
\ee
and the potential 
\ba
V_{{\rm het}} & =& {1 \over 32} e^{2 \chi} \Big(
\cosh 2\phi (\cosh 2\om_+ \cosh 2\om_- -1)
\nonumber\\
&& - 2 ( \sinh^2 \om_+ +  \sinh^2 \om_- + 4 \sinh\om_+ \sinh\om_-)\Big) \ .  
\label{h9j}
\ea
The kinetic term for the fields $\phi$, $\chi$ and $\om_\pm$ is as in 
\eqn{llkin}.
The truncated system of differential equations is 
\ba
{d \chi \over dr} & = & -{1 \over 2\sqrt{2}} e^{\chi} 
\left( e^{\phi} - 
\sinh \phi \sinh\omega_+ \sinh \om_- \right)\ , 
\nonumber \\
{d \phi \over dr} & = & -{1 \over \sqrt{2}} e^{\chi} \left(
e^{\phi} - \cosh\phi \sinh\omega_+ \sinh \om_-\right) \ , 
\label{het3}\\
{d \omega_\pm \over dr} & = & {1\ov \sqrt{2}} e^{\chi} 
{\rm sinh}\phi \cosh \om_\pm \sinh\om_\mp ~. 
\nonumber 
\ea
The equation for the 
conformal factor of the metric \eqn{metriki} is easily integrated and gives 
$A=-\chi$, as in the case of the genuine type-IIA or type-IIB theories.
Also, as in the previous cases the winding fields $\om_\pm$ are
related by \eqn{lala} and we may take the constant $\l\geq 1$ 
with no loss of generality.
Similar to before,
for $\om_+=\om_-$ we obtain the corresponding expressions in section 5
of \cite{BBDS}.

\medskip
\no
\underline{The critical points}:  These occur in the present case 
only when $\om_+ \om_- > 0$ and they are located at 
\be
\phi = 0 ~, \qq \omega_{+} = \ln \left(\sqrt{1+1/\l^2}+1/\l\right)\ ,\qq
\omega_{-} = \ln \left(\sqrt{1+\l^2}+\l\right)\ 
\label{crii}
\ee
and at the mirror image point that appears
because of the invariance of the equations 
under $\omega_\pm \rightarrow -\omega_\pm$. These points are
minima of the potential \eqn{h9j} which, when evaluated for either 
of these points, 
becomes $V=-{1\ov 8} e^{2\chi}$ and is independent of the constant $\l$.
The dilaton equation in \eqn{het3} is easily integrated to give 
at these critical points $e^{-\chi} \sim r$.
Recall now that the string frame metric in four dimensions 
is obtained by \eqn{metriki} after 
multiplying with $e^{2 \chi}$. Hence, the resulting background has a 
flat string metric and a linear dilaton (in the coordinate 
$\zeta \sim \ln r$).\footnote{Note that, in this case, 
it makes sense to pass from the Einstein 
to the string frame since,
when $\phi$ and $\om$ assume their critical values \eqn{crii},
we are left with only massless fields (graviton and dilaton) that 
couple to a 2-dimensional string world-sheet action.} This is an exact 
solution in 
string theory. For the case with $\l=1$, corresponding to 
$\om_+=\om_-=\ln(\sqrt{2}+1)$
this solution was found, in the presence context, in \cite{ak,adk},
by solving directly the second order equations of motion following 
from the action.
What we have shown here is that the solution persists for more general values
for the windings, parametrized by a constant $\l$ as in \eqn{crii}.
Finally, we notice that these critical points are saddle points, as
a stability analysis around them reveals.

\medskip
\no
\underline{The general solution}: The general solution of the system of eqs. 
\eqn{het3} cannot be found in a closed form. 
The case with $\om_-=\om_+$ was thoroughly investigated in \cite{BBDS}
and it has a rather complicated space of solutions (see 
fig. 3 of \cite{BBDS}). 
We expect that the same is true in the more general case with $\om_+\om_->0$.
In contrast, when $\om_+\om_-<0$, it is more or less
straightforward to show that a general solution in the $\phi$--$\om$ plane
starts in the region with $(\phi,\om)=(-\infty, +\infty)$ and ends in the 
region with $(\phi,\om)=(+\infty, +\infty)$. In between, 
$\om$ acquires a minimum
value $\om_{\rm min}$ that can be any positive
number, depending on the constants of integration.
At the two ends we have strong coupling regimes. In particular,
\be
e^{-\om} \sim e^{-\phi}\sim e^{-2\chi}\sim r^{2/7}\ ,
\qq {\rm as} \quad r\to 0^+\ 
\label{strr1}
\ee 
and similarly, 
\be
e^{-\om} \sim e^{\phi}\sim e^{-2\chi}\sim (r_0-r)^{2/7}
\ ,\qq {\rm as} \quad r\to r_0^-\ .
\label{strr2}
\ee
In both cases the metric takes the form \eqn{hghg3} with $\n=1/7$.
According to our criterion this solution is `unphysical'.

\medskip
\no
\underline{The truncation in the limit $\om_-\to\pm \infty$}: Consider 
the rescaling
followed by the limit
\be
\chi \to \chi -\ln \l\ ,\qq \l\to \infty\ .
\ee
Then, after substituting $\sinh\om_-\simeq \l \cosh \om$ using \eqn{lala},
the system of eqs. \eqn{het3} simplifies to
\ba
{d \chi \over dr} & = & \mp{1 \over 4\sqrt{2}} e^{\chi} 
\sinh \phi \sinh 2\om \ , 
\nonumber \\
{d \phi \over dr} & = & \mp {1 \over 2 \sqrt{2}} e^{\chi} \cosh\phi 
\sinh 2\omega \ , 
\label{het53}\\
{d \omega\over dr} & = & \mp {1\ov \sqrt{2}} e^{\chi} 
{\rm sinh}\phi \cosh^2 \om ~,
\nonumber 
\ea
where the upper (lower) sign corresponds to the case with $\om_+ \om_- >0$ 
($\om_+ \om_- <0$).
The general solution of this system is given in parametric form by
\be
\cosh \phi =(\const) \cosh\om \ ,\qq e^{2\chi} = \cosh\om\ ,
\ee
where the plus and minus signs are unrelated to the ones in \eqn{het53}.
It is easily seen that the strong coupling regions exhibits the general 
behaviour of \eqn{strr1} and \eqn{strr2}.

\section{Concluding remarks and further comments}

In this paper we pursued further the idea, put forward in \cite{BBDS},
that superstrings at finite temperature may reach stable phases and 
therefore they do not
necessarily suffer from Hagedorn-type instabilities. Our investigation was 
based on an effective supergravity theory
that takes into account the relevant winding states 
of the string and the interactions among themselves and with the massless 
moduli \cite{ak,adk}. First we constructed the most general truncation
in the real-field sector of these theories and formulated general
conditions for the existence of 1/2-BPS states in this theory.
These take the 
form of 9 coupled, first order non-linear differential equations.
In the particular cases of the 
type-IIA, type-IIB and a self-dual hybrid, theories,
we presented the most general solutions
in closed form and investigated their properties. For the heterotic case
the general solution cannot be found in closed form.
Nevertheless we analyzed its properties using the system of 
differential equations that it obeys and  
presented it in closed form for a particular case.

The solutions that we presented in this paper, compared to the similar ones
in \cite{BBDS}, allow for unequal left and right winding fields. They also
have the important feature that flat space with constant moduli 
is recovered in 
the limit of small right-winding field. In addition,
we have presented a mechanism
for changing the sign of the winding field, in an effective action approach,
by requiring smooth continuations of our solutions through a curvature 
singularity.

An issue concerning the supersymmetric solutions found in
this paper and also in \cite{BBDS}, is whether they suffer from Jens 
instabilities, typical in thermodynamical systems that contain gravity.
A prototype such example was studied in \cite{gross} where
hot flat space was found to be unstable under long-wavelength density 
fluctuations of thermal gravitons (expected from the classical Jens 
instability) as well as due to nucleation of black holes (quantum 
instability). 
These instabilities render the thermal canonical ensemble ill defined. 
In turn, one may wonder if the very foundations of the theories \cite{ak,adk}
should be doubted in the sense that the identification of the relevant fields
$\{S,T,U,Z^\pm_A\}$ was based on perturbative string compactifications
that were interpreted as thermal ensembles.  
One is tempted to conclude that such 
instabilities cannot be avoided in our solutions especially due to the fact 
that they do not support small volumes and according to general arguments
they should collapse into black holes \cite{aw}.
However, one may present the counter argument that the supersymmetric 
property of our solutions will ensure their quantum stability as well and that 
no gravitational collapse will occur.
We note that the analysis of \cite{gross} was done around hot flat space
and supersymmetry was not even an issue. Our spaces have very different 
asymptotic behaviours and we think that the conclusions of \cite{gross} are
not directly applicable to our cases.

A perhaps related question is whether an effective action approach 
can capture a possible breaking of supersymmetry by thermal effects at 
arbitrarily high mass levels. This is an open question, but it should be noted
that higher massive modes can be included in an effective action approach,
though technically this is certainly an involved construction.
We believe that the inclusion of more states enhances rather than diminishes 
the chances that supersymmetry can be preserved since it opens up to more
possibilities.

Another important issue concerning the supersymmetric solutions found in
this paper, as well as those in \cite{BBDS}, is to understand their microscopic
origin. For that, a `lift' to string or M-theory is necessary. Such a `lift'
will also elucidate the origin of the naked
singularities that our solutions have. 
Similar studies have appeared in the recent
literature in relation to solutions of five-, four- and 
seven-dimensional gauged 
supergravities and the corresponding `lifted' solutions in 
string or M-theory. 
Without entering into details we mention that,
naked singularities in the lower 
dimensional theories sometimes 
appear benign and have natural interpretations
from a higher dimensional point of view.
We expect that at least a subset of our solutions
that were characterized as `physical' according 
to our semi-classical criterion,
will survive a real stringy test 
(see also the remarks after eq. \eqn{hghg3}). 
We hope that work along these lines will
be reported in the future.

\bigskip\bigskip
\centerline{\bf Acknowledgements}
\noindent
I would like to thank the theory division at CERN for hospitality 
and financial support during a considerable part of this work.
This research was supported
by the European Union under contracts 
TMR-ERBFMRX-CT96-0045 and -0090, by the Swiss Office for Education and
Science, by the Swiss National Foundation and by RTN contract 
HPRN-CT-2000-00122.


\end{document}